\documentclass[twocolumn,showpacs,preprintnumbers,amsmath,amssymb]{revtex4}

\usepackage{epsf}
\usepackage{dcolumn}
\usepackage{bm}

\newcommand{\bea}{\begin{eqnarray}}
\newcommand{\eea}{\end{eqnarray}}
\newcommand{\ba}{\begin{array}}
\newcommand{\ea}{\end{array}}

\begin{document}

\title{Rapid Ballistic Readout for Flux Qubits}

\author{Dmitri V. Averin, Kristian Rabenstein, and Vasili K. Semenov}

\affiliation{Department of Physics and Astronomy, Stony Brook
University, SUNY, Stony Brook, NY 11794-3800}

\date{\today}

\begin{abstract}
We suggest a new type of the magnetic flux detector
which can be optimized with respect to the measurement
back-action, e.g. for the situation of quantum  measurements. The
detector is based on manipulation of ballistic motion of
individual fluxons in a Josephson transmission line (JTL), with
the output information contained in either probabilities of fluxon
transmission/reflection, or time delay associated with the fluxon
propagation through the JTL. We calculate the detector
characteristics of the JTL and derive equations for conditional
evolution of the measured system both in the
transmission/reflection and the time-delay regimes. Combination of
the quantum-limited detection with control over individual fluxons
should make the JTL detector suitable for implementation of
non-trivial quantum measurement strategies, including conditional
measurements and feedback control schemes.

\end{abstract}

\pacs{  }

\maketitle

\section{Introduction}

One direction of current efforts aimed at development of
mesoscopic solid-state qubits is realization of effective schemes
of qubit measurements. Different versions of the qubits which use
quantum dynamics of magnetic flux in superconducting loops
\cite{b1,b2,b3,b4,b5,b6,b7,b8,b9} are at the moment among the most
advanced solid-state qubits. Practically all of them employ the
measurement scheme based on the modulation by the measured qubit
of the decay rate of the supercurrent of a Josephson junction or a
SQUID. This process can be viewed as tunneling of a magnetic flux
quantum and has several attractive features as the basis for
measurement. Most important one is sufficiently large sensitivity
which comes from strong dependence of the tunneling amplitude on
parameters of the tunneling potential controlled by the measured
system. There is, however, an important disadvantage of this
approach. In simple few-junction structures, the supercurrent
decay brings the detector into the finite-voltage state
characterized by large energy dissipation, which strongly perturbs
both the system and the detector itself, and makes it impossible
to repeat the measurements sufficiently quickly. This strong
perturbation also destroys the state of the measured quantum
system by changing it in an uncontrolled way. Both of these
factors prevent realization of non-trivial quantum measurement
strategies which are typically based on continuous measurements or
a sequence of successive discrete measurements. The goal of this
work is to suggest and analyze a new flux detector which should
not have this drawback. While still based on tunneling of
individual magnetic flux quanta, the detector nevertheless avoids
the transition into the dissipative state by using ballistic
motion of the flux quanta \cite{rem1}. The detector should have
the quantum-limited back-action on the measured system and time
resolution sufficient for realization of non-trivial quantum
measurements of superconducting qubits. The cost of achieving this
is the need for single-flux-quantum (SFQ) support electronics
\cite{b10} required for operation of this detector. Adaptation of
the SFQ circuits to qubit applications \cite{b11,b12*,b13*} is an
important, and not fully solved, problem of development of
scalable superconducting qubits.

\section{Josephson Transmission Line as Flux Detector}

The suggested detector is based on the ballistic motion of fluxons
in the Josephson transmission line (JTL) formed by unshunted
junctions with critical currents $I_C$ and capacitances $C$
coupled by inductances $L$ as shown in Fig.~1. The detector can be
viewed as the flux analog of the quantum point contact (QPC)
charge detectors (see \cite{b12} and references therein) used for
measurements of the quantum-dot qubits. Both detectors make use of
the ability of the measured system to control the ballistic motion
of independent particles (electrons in the QPC, fluxons in the JTL
detector) through a one-dimensional channel. The JTL detector
should, however, provide more control over the propagation of
individual fluxons than is possible with electrons in the QPCs.
This leads to new measurement regimes (e.g., time-delay
measurement) impossible with the QPC.

\begin{figure}[h]
\setlength{\unitlength}{1.0in}
\begin{picture}(3.3,1.6)
\put(.0,-.1){\epsfxsize3.3in\epsfbox{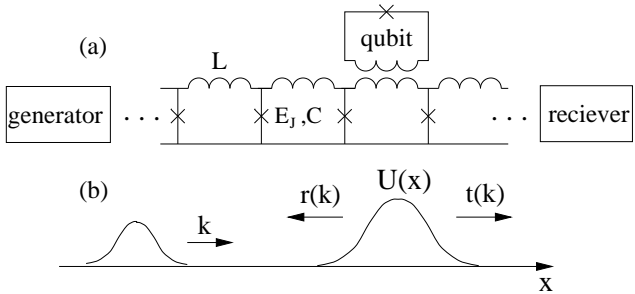}}
\end{picture}
\caption{(a) Equivalent circuit of the flux detector based on the
Josephson transmission line (JTL) and (b) diagram of scattering of
the fluxon injected into the JTL with momentum $k$ by the
potential $U(x)$ that is controlled by the measured qubit. The
fluxons are periodically injected into the JTL by the generator
and their scattering characteristics (transmission and reflection
coefficients $t(k)$, $r(k)$) are registered by the receiver.}
\end{figure}

The JTL detector uses the fact that the flux $\Phi^{(e)}(x)$
generated by the measured system creates potential $U(x)$ for the
fluxons moving in the JTL (Fig.~1). The fluxons are injected one
by one, with the period $1/f$, into the JTL by the generator, and
are scattered by the potential $U(x)$ localized in some region of
the JTL. The injection frequency $f$ is sufficiently low so that
only one fluxon at a time moves inside the JTL. The fluxon
scattering characteristics: transmission probability or the time
delay associated with the motion through the JTL, are registered
by the receiver. Since these characteristics depend on the
potential $U(x)$ controlled by the measured system, they contain
information about the state of this system.

Quantitatively, Hamiltonian of the uniform JTL can be expressed in
terms of the electric charges $Q_n$ and the phases $\phi_n$ of the
Cooper-pair condensate at each of its nodes $n$:
\begin{eqnarray}
H=\sum_{n}\big[Q_n^2/2C+E_J(1-\cos \phi_n)  \nonumber \\+E_L
(\phi_{n+1} -\phi_n -\phi_n^{(e)})^2 \big] \, . \label{e1}
\end{eqnarray}
In this Hamiltonian, $E_J=\hbar I_C/2e$ is the Josephson coupling
energy of the junctions, where $I_C$ is the junction critical
current, $E_L=(\Phi_0/2 \pi )^2/2L$ is the characteristic magnetic
energy of inductances $L$, where $\Phi_0=\pi \hbar/e$ is the
magnetic flux quantum, and $\phi_n^{(e)}=2\pi\Phi_n^{(e)}/\Phi_0$
is the phase difference across the inductance $L$ of the $n$th
segment induced by the external magnetic flux $\Phi_n^{(e)}$
through this segment. The charges and phases at each node are the
conjugate variables that satisfy canonical commutation relations:
$[\phi_n,Q_m]=2ei \delta_{nm}$.

In what follows, we are interested in the regime of small
inductances, $L \ll \Phi_0/I_C$, in which the JTL acts as a
uniform ballistic channel for the fluxons. The phase difference
across each segment of the JTL is then small, and one can replace
the phases $\phi_n$ with continuous function $\phi(x)$ of
dimensionless coordinate $x=n$ along the JTL. (Condition of
validity of this approximation is discussed more quantitatively at
the end of this Section.) In this regime, the JTL is equivalent to
a distributed Josephson junction, and the Hamiltonian (\ref{e1})
reduces to the standard ``sine-Gordon'' Hamiltonian, which can be
written as follows:
\begin{eqnarray}
H=2E_L \int dx \Big\{ \frac{1}{2}[(\partial_{\tau}\phi)^2+
(\partial_{x}\phi)^2] +  \nonumber \\ + \lambda_J^{-2}(1-\cos
\phi) - \phi^{(e)}(x) \partial_{x}\phi \Big\}\, . \label{e2}
\end{eqnarray}
Here $\tau \equiv ct$ is the time $t$ normalized to the velocity
$c=1/\sqrt{LC}$ of propagation of excitations along the JTL in the
absence of Josephson tunneling, and $\lambda_J=(\hbar/2e I_C
L)^{1/2}$ is the Josephson penetration length of the junction. We
note that in the notations used in Eq.~(\ref{e2}) and below, all
distances along the JTL, including $x$ and $\lambda_J$, are
dimensionless, and are measured in units of the cell size $a$ of
the JTL (see Fig.~3). Conversion of most equations to absolute
units of length can be achieved simply by changing interpretation
of all quantities (e.g., $L,\, C,\,I_C$) from ``per cell'' to
``per unit length''.

The commutation relations for the charges and phases give the
following equal-time commutation relations for the field
$\phi(x,\tau)$ in the Hamiltonian (\ref{e2}):
\begin{equation}
[\phi(x),\partial_{\tau}\phi(x')]= \beta^2 \delta (x-x') \, ,
\label{e3}
\end{equation}
where the parameter $\beta^2\equiv (4e^2/\hbar) \sqrt{L/C}$
measures the wave resistance $\sqrt{L/C}$ of the JTL in the
absence of Josephson tunneling relative to the quantum resistance.
Known results for the quantum sine-Gordon model (see, e.g.,
\cite{sol}) show that when $\beta^2\geq 8\pi$, i.e.,
$\sqrt{L/C}\geq h/e^2 \simeq 25$ k$\Omega$, quantum fluctuations
of the field $\phi$ completely destroy the quasiclassical
excitations of the junction. The transition at $\beta^2= 8\pi$
should be qualitatively similar to the analogous resistance-driven
transition in small Josephson junctions \cite{b13}. This analogy
suggests that the dynamics of the supercurrent flow in the JTL
with large wave resistance should be described in terms of
tunneling of individual Cooper pairs \cite{b14}. While this limit
might be reachable in very narrow and thin JTLs of sub-micron
width \cite{b15}, we assume here more typical situation, when
$\sqrt{L/C}$ is on the order of 10 -- 100 Ohm and $\beta^2 \ll 1$.
In this case, the JTL supports a number of quasiclassical
excitations including, most importantly for this work, topological
solitons that carry precisely one quantum of magnetic flux each.
Dynamics of such ``fluxons'' is equivalent to that of stable, in
general relativistic, particles \cite{sol} with the terminal
velocity $c=1/\sqrt{LC}$ and mass
\begin{equation}
m\simeq 8\hbar \omega_p/c^2\beta^2 = (2\hbar/e)^{3/2} (I_C L
)^{1/2} C \, , \label{e4}
\end{equation}
where $\omega_p= (2eI_C/\hbar C)^{1/2}=c/\lambda_J$ is the plasma
frequency. Another type of quasiclassical excitations in the JTL
are the small-amplitude plasmon waves with frequency
\begin{equation}
\omega (k) = (\omega_p^2+c^2k^2)^{1/2} \, , \label{e5}
\end{equation}
for a wavevector $k$.

In this work, we are interested in the ``non-relativistic'' regime
of fluxon dynamics, when velocity $u$ of its motion is small,
$u\ll c$. Equations (\ref{e4}) and (\ref{e5}) show that in this
regime, the fluxon kinetic energy $\epsilon=\hbar^2 k^2/2m$, where
$k=mu$, can be made smaller than the lowest plasmon energy $\hbar
\omega_p$,
\begin{equation}
\epsilon = \hbar \omega_p \cdot (2u/c\beta)^2 \, , \label{e51}
\end{equation}
so that for $u<c\beta/2$ the fluxon can not emit a plasmon even
when it is scattered by non-uniformities of the JTL potential
\cite{b16}. Intrinsic dissipation associated with emission of
plasmons is then suppressed, and the fluxon motion in the JTL
should be elastic, provided that other, ``extrinsic'', sources of
dissipation are also sufficiently weak. Although the JTL operation
as the flux detector should be possible for moderately strong
fluxon dissipation, the dissipation would prevent the detector
from reaching the quantum-limited regime.

The shape of the scattering potential $U(x)$ for fluxons created
by the measured system is determined by the convolution of the
distribution of the flux $\Phi^{(e)}(x)$ with the distribution of
current in each fluxon \cite{p1,p2}, and can be written as:
\begin{equation}
U(x)= \frac{\Phi_0}{2 \pi L} \int dx' \frac{\partial
\Phi^{(e)}(x')}{\partial x'} \phi_0(x'-x) \, , \label{e6}
\end{equation}
where $\phi_0(x)$ is the shape of the fluxon that in general can
be distorted by the potential $U(x)$ itself. If, however, the
potential is smaller than $\omega_p$, or does not vary appreciably
on the scale of the size of the fluxon given by $\lambda_J$, the
changes in the fluxon shape are negligible, and one can use in
Eq.~(\ref{e6}) the regular fluxon shape in the uniform case, which
in the non-relativistic limit is $\phi_0(x)= 4 \tan^{-1} [\exp(x/
\lambda_J )]$. One of the implications of Eq.~(\ref{e6}) is that
the width of the scattering potential $U(x)$ can not be made
smaller than $\lambda_J$.

If the measured system is coupled to the JTL not inductively (as
in Fig.~1) but galvanically, and injects the current $j^{(e)}(x)$
in the nodes of the JTL, potential created by this current for the
fluxons is still given by Eq.~(\ref{e6}) if one makes the
substitution
\begin{equation}
(1/L) \partial \Phi^{(e)}(x)/\partial x= j^{(e)}(x)\, . \label{e7}
\end{equation}
Discrete version of Eq.~(\ref{e7}) is illustrated in Fig.~2.
External flux $\Phi^{(e)}_n$ through the $n$th cell of the JTL
induces circulating current $j_n=\Phi^{(e)}_n/L$ in this cell. The
difference between the currents $j_n$ in the neighboring cells
creates the currents $j^{(e)}_n$ through the Josephson junctions
of the JTL: $j^{(e)}_n=j_n-j_{n-1}$. In the continuous limit, this
relation gives Eq.~(\ref{e7}). It implies that direct injection of
the currents $j^{(e)}_n$ in the JTL junctions indeed produces the
same distribution of phases and currents as magnetic coupling
generating the fluxes $\Phi^{(e)}_n$. Throughout this work we
assume that the measured system does not inject the net current in
the JTL, $\int dx j^{(e)}(x)=0$, i.e., the corresponding flux
(\ref{e7}) and the potential (\ref{e6}) vanish outside the
coupling region: $\Phi^{(e)}(x) \rightarrow 0$ for $x \rightarrow
\pm \infty$.

\begin{figure}[h]
\setlength{\unitlength}{1.0in}
\begin{picture}(3.3,1.0)
\put(.6,-.05){\epsfxsize2.1in\epsfbox{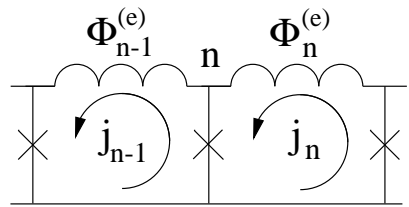}}
\end{picture}
\caption{Equivalence between the magnetic and galvanic coupling to
the detector. External fluxes $\Phi^{(e)}_n$ through the JTL cells
induce the circulating currents $j_n$ in them which in their turn
create the currents $j^{(e)}_n=j_n-j_{n-1}$ through the JTL
junctions. The situation would be the same if the currents
$j^{(e)}_n$ are injected directly into the JTL junctions by
external system.}
\end{figure}

Although the assumed condition $\beta^2 \ll 1$ makes quantum
fluctuations of the fluxon shape small, the dynamics of the fluxon
as a whole can still be completely quantum. An example of quantum
dynamics of fluxons of this type was recently observed
experimentally \cite{tun}. JTL detector uses the quantum fluxon
dynamics in the potential described by Eqs.~(\ref{e6}) or
(\ref{e7}). As was mentioned above, operation of the detector
requires that the fluxons are injected by the generator into one
end of the JTL and observed by the receiver at the other end
(Fig.~1). Both of these circuits can be designed following the
general principles of the SFQ electronics \cite{b10,b12*} and will
not be discussed explicitly here. We simply assume that the ends
of the JTL are matched appropriately to these circuits so that the
fluxons can enter and leave the JTL without reflection and are
injected in the JTL in an appropriate quantum state. This initial
state $\psi(x,t=0)$ of an injected fluxon is characterized by the
average fluxon velocity $u$ and the wave-packet $\psi_0(x)$
defining its position:
\begin{equation}
\psi(x,t=0) = \psi_0(x)e^{ik_0 x} \, , \;\; k_0 = mu. \label{e8}
\end{equation}
As we will see from the discussion in the next Section, many
properties of the JTL detector are independent of the specific
shape of the wave-packet $\psi_0(x)$, as long as it is well
localized both in coordinate space and space of the fluxon
momentum $k$. They depend only on the wave-packet width $\xi$ in
coordinate space and the corresponding width $\delta k \simeq
1/\xi$ in momentum space. These parameters should satisfy two
obvious conditions: $\xi \ll l$, where $l$ is the total length of
the JTL, and $\delta k \ll k_0$. We assume a stronger form of the
second condition that follows from the requirement that the
broadening of the wave-packet by $\delta x \simeq \delta k t/m$
because of the momentum uncertainty $\delta k$ during the typical
time $t\simeq l/u$ of the fluxon propagation through the JTL is
negligible in comparison to the initial width: $\delta x\ll \xi $.
The discussion below shows that this requirement is necessary for
the quantum-limited operation of the JTL detector in the
time-delay mode. The two conditions mean that the width $\xi$ is
limited as follows:
\begin{equation}
(l/k_0)^{1/2} \ll \xi \ll l \, . \label{e9}
\end{equation}

In cases when it will be necessary to specify the shape of the
wave-packet $\psi_0(x)$ we will take it to be Gaussian:
\begin{equation}
\psi_0 (x) = (\pi \xi^2)^{-1/4} e^{-(x-\bar{x})^2/2\xi^2}\, ,
\label{e10}
\end{equation}
where $\bar{x}$ is the initial fluxon position in the JTL. Besides
being well localized as necessary both in the momentum and
coordinate space, the wave-packet (\ref{e9}) can be obtained as a
result of the fluxon generation process that can be implemented
naturally with the SFQ circuits. This process consists of the two
steps: relaxation of the fluxon to the ground state of a weakly
damped and nearly quadratic Josephson potential with the required
width $\xi$ of the wavefunction of the ground state, and then
rapid acceleration to velocity $u$ after this potential is
switched off.

While the assumed condition $\beta^2 \ll 1$ does not preclude the
quantum dynamics of fluxons in the JTL, Eq.~(\ref{e4}) shows that
decreasing $\beta$ increases the fluxon mass. This makes it more
difficult to maintain quantum coherent dynamics of fluxons, which
in the case of large mass becomes more susceptible to
perturbations. For instance, to avoid the effects  of discreteness
on the fluxon dynamics in the case of the discrete JTL structure
(Fig.~1), the wavelength of its wavefunction (\ref{e8}) should be
larger than the size of one cell. This condition can be written as
$k_0 \lesssim 1$ and means that the fluxon velocity and kinetic
energy can not be larger than, respectively $u=\hbar /m = c \cdot
(\beta^2 \lambda_J/8)$ and $\epsilon =\hbar^2 /2m$, i.e.
\begin{equation}
\epsilon = \hbar c \beta^2 \lambda_J/16=(e^2/C)(\lambda_J/4) \, .
\label{e52}
\end{equation}
We see that both the fluxon velocity $u$ and its kinetic energy
$\epsilon$ decrease with decreasing $\beta^2$ and any external
perturbation (e.g., fluctuations of the critical current of the
JTL junctions) affects the fluxon motion more strongly at small
$\beta$. For realistic values of $\lambda_J$ and $\beta$ the
limitation on the fluxon energy (\ref{e52}) is stronger than
(\ref{e51}). Since the fluxon size $\lambda_J$ can not be
increased much for obvious reasons of convenience of fluxon
manipulation, the energy  (\ref{e52}) depends only on the junction
capacitance $C$, and can be increased only by decreasing the
junction size (see the discussion below). In practical terms, the
fluxon energy (\ref{e52}) should be at least larger than
temperature for the fluxons to be generated by an SFQ generator.
All this means that the JTL can operate as the quantum detector
only with junctions of sub-micron size. Although the estimate
(\ref{e52}) was made for the discrete JTL, the same practical
conclusion would be reached in the case of nominally-uniform long
Josephson junctions, for which the limitation on the fluxon
momentum $k_0$ would be set by unavoidable spatial fluctuations of
the junction parameters.

\begin{figure}[h]
\setlength{\unitlength}{1.0in}
\begin{picture}(3.3,1.35)
\put(.4,-.1){\epsfxsize2.1in\epsfbox{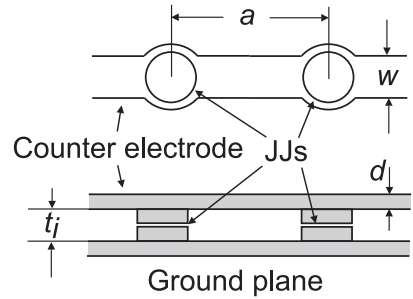}}
\end{picture}
\caption{Geometry of one cell of the multi-layer JTL: top view and
vertical cross-section. Internal layers of the structure are used
to create Josephson junctions (JJs), while the two external
layers, the ground plane and the counter-electrode, separated
vertically by the distance $t_i$, define inductance $L$. The
length $a$ of the cell is set by the distance between the
junctions.}
\end{figure}

We end this Section with a discussion of experimental realization
of the JTL detector with conventional thin-film ``tri-layer''
fabrication technology of the SFQ circuits. Simplified but
sufficiently realistic geometry of one JTL cell in this case is
shown in Fig.~3. It contains 4 superconductor (niobium) films, the
lower and the top of which (ground plane and the
counter-electrode) are used to define inductances, while the two
layers in the middle are separated by a very thin layer of
insulator (the tri-layer structure) and are used to produce the
Josephson junctions. Such a discrete multi-layer configuration of
the JTL is more convenient than a uniform long Josephson junction,
since it enables one to optimize separately the junctions and
inductances of the JTL, most importantly, to reduce effective
capacitance.

One possible drawback of the discrete configuration could be the
parasitic periodic potential $u(z)$ for the fluxons produced by
the JTL discreteness (see, e.g., \cite{dis}). This potential,
however, is negligible when $\lambda_J$ is not too small,
$\lambda_J \geq 2$. Indeed, using the Hamilonian (\ref{e1}) and
equation for the static distribution $\phi_n$ that follows from
it, we can write the energy $E$ of the static fluxon as
\begin{equation}
E=E_J\sum_{n} \epsilon_n \, ,\;\;\; \epsilon_n \equiv 1-\cos
\phi_n-\frac{1}{2} \phi_n \sin \phi_n  \, . \label{e53}
\end{equation}
Applying the Poisson summation formula to this expression, we get
in the limit of sufficiently large $\lambda_J$:
\begin{equation}
E=E_0[1+\alpha \cos (2\pi z)] \, , \label{e54}
\end{equation}
where $E_0 = mc^2$ is the fluxon rest energy, $z$ is the fluxon
coordinate in units of the JTL period $a$, and
\begin{equation}
\alpha = \frac{1}{8}\int dy \epsilon (y) e^{i 2\pi y \lambda_J} \,
. \label{e55}
\end{equation}
Here $\epsilon (y)$ is a smooth envelope of the discrete
distribution $\epsilon_n$ (\ref{e53}) associated with the static
fluxon $\phi_n$. Since $\epsilon (y)$ is smooth in the limit of
large $\lambda_J$, Eq.~(\ref{e55}) implies that the amplitude
$\delta \epsilon $ of the $z$-dependent part of the fluxon energy
is exponentially small in $\lambda_J$. Specifically, if $\epsilon
(y)$ is taken as for the regular ``continuous'' fluxon $\phi(y)= 4
\tan^{-1}[e^y]$, we see that
\[ \alpha \propto e^{-\pi^2 \lambda_J}, \]
and periodic potential associated with the JTL discreteness is
completely negligible for $\lambda_J \geq 2$ even on the scale of
small fluxon kinetic energy (\ref{e52}). For operation of the JTL
detector it is obviously convenient to keep $\lambda_J$ (i.e., the
fluxon size) as close as possible to a minimum consistent with
this inequality.

In the multi-layer fabrication technology (Fig.~3), the Josephson
junctions have the critical current density $j_{c}$ that can be
varied within wide range, $10 \div 10^{4}$ A/cm$^{2}$, while
specific junction capacitance $c_{J}$ changes only a little, from
30 fF/$\mu $m$^{2}$ to 60 fF/$\mu $m$^{2}$, in this range of
$j_{c}$'s. This means that there is a strong limitations on the
minimal junction capacitance $C$ set by the junction diameter $D$,
e.g., $C\simeq 30$ fF for $D\simeq 1.0\, \mu $m. Inductances in
the multi-layer JTL (Fig.~3) are characterized by the specific
inductance per square $\kappa=\mu_0\cdot (t_{i}+2\lambda )$, where
$\lambda$ is the superconductor penetration depth, and parasitic
capacitance $c_{L}\approx \epsilon \epsilon_0 /t_i$ per unit area.
For realistic $t_i\approx 0.2\div 0.4$ $\mu $m, $\kappa$ and
$c_{L}$ are $0.5\div 0.7$ pH and $0.2\div 0.1$ fF/$\mu $m$^{2}$,
respectively.

The discussion leading to Eq.~(\ref{e52}) shows that the main
requirement on the JTL parameters optimizing its operation as the
quantum-limited detector consists in making the junction
capacitance $C$ as small as possible. After this, the inductance
$L$ can also be reduced as long as this reduction is at least
partly compensated by increase in $j_c$ and does not make the
characteristic fluxon size $\lambda_J$ too large. The actual
fluxon size is about 4 $\lambda_J$, and since it is impractical to
have significantly more than about 30 cells in the JTL,
$\lambda_J$ is basically fixed within the range from 2 to 3. For
relatively small junction size $D= 0.3\, \mu $m and high critical
current density $j_{c}=10$ kA/cm$^{2}$, the value of $\lambda_J$
within this range corresponds to $L \simeq 4$ pH, inductance that
is obtained for the width of the superconductor strip $w=1\,
\mu$m, and the cell size $a \simeq 5\, \mu$m. For these
parameters, the parasitic capacitance $c_L$ gives noticeable
contribution (about 30\%) to the effective junction capacitance $C
\simeq 4$ fF. The propagation velocity along such a JTL is
$c\simeq 10^{13}$ cells/s, i.e., $5\cdot 10^7$ m/s, and the wave
resistance is about 30 Ohm giving $\beta^2 \simeq 0.03$. Useful
fluxon velocity in the quantum-limited regime [estimated from
Eq.~(\ref{e52})] is $10^{11}$ cells/s, so that the total time of
the fluxon propagation through the JTL (which limits the detector
time resolution) is about 0.3 ns. We se that the JTL detector
should have very good time resolution even in the quantum-limited
regime, but requires in this case the JTL parameters which are
close to the limit of the current SFQ fabrication technology.

If one abandons the goal of realizing the quantum-limited
detection, requirements on the JTL parameters become much more
routine. A typical set of parameters $D \simeq 1.5\, \mu $m,
$j_{c}=1$ kA/cm$^{2}$, and $w=3\, \mu$m, $a=10\, \mu$m, produces
the JTL with $c\simeq 3 \cdot 10^{12}$ cells/s, wave resistance
about 2 Ohm, and $\lambda_J \simeq 2$. This should give the
detector that is not quantum-limited but is very fast and has the
time resolution on the order of 0.1 ns. As discussed in Sec.~5,
such a detector could be used in non-trivial schemes of quantum
measurement despite classical dynamics of fluxons in it.

\section{Measurement dynamics of the JTL detector}

As discussed above, the measurement by the JTL detector consists
qualitatively in scattering of the fluxons in the JTL by the
potential controlled by the measured system. This process has the
simplest dynamics if the measured system is stationary. In this
case it is convenient to consider it in the basis of the
eigenstates $|j\rangle$ of the system operator (e.g., magnetic
flux in the qubit loop in the example shown in Fig. 1) which
couples it to the JTL. In each state $|j\rangle$, the system
creates different potential $U_j(x)$ for the fluxons propagating
through the JTL. Different realizations $U_j(x)$ of the JTL
potential produce different scattering coefficients for injected
fluxons: the amplitude $t_j(k)$ of transmission through the JTL,
and the amplitude $r_j(k)$ for the fluxon to be reflected back
into the generator. Both amplitudes are in general functions of
the fluxon momentum $k$. Since they depend on the state
$|j\rangle$ of the measured system, scattered fluxons carry
information about $|j\rangle$.

In general, the process of quantum measurement can be understood
as creation of entangled state between the measured system and
detector as a result of interaction between them. The states of
the detector are quasiclassical and suppress quantum superposition
of different outcomes of measurement. The two consequencies of
this process are the acquisition of information about the system
by the detector and ``back-action'' dephasing of the measured
system - see e.g., \cite{m1}. For the JTL detector, the
detector-system entanglement arises as a result of fluxon
scattering, and the rates of information acquisition and
back-action dephasing can be expressed in terms of the scattering
parameters. In this respect, the JTL is very similar to the QPC
detector and its characteristics can be obtained following the
derivation \cite{b12} for the QPC.

Evolution of the density matrix $\rho$ of the measured system in
scattering of one fluxon can be obtained by considering first the
time dependence of the total wavefunction of the fluxon injected
into the JTL and the wavefunction $\sum_j c_j |j\rangle$ of the
measured system:
\begin{equation}
\psi(x,t=0) \cdot \sum_j c_j|j\rangle \rightarrow \sum_j
c_j\,\psi_j(x,t) \cdot |j\rangle \, . \label{e11}
\end{equation}
Here the initial fluxon wavefunction $\psi(x,t=0)$ is given by
Eq.~(\ref{e8}) and its time evolution $\psi_j(x,t)$ depends on the
realization $U_j(x)$ of the scattering potential created by the
measured system. Evolution of $\rho$ is obtained then by tracing
out the fluxon part of the wavefunctions (\ref{e11}):
\begin{equation}
\rho_{ij} =c_ic_j^* \rightarrow c_ic_j^* \int dx \psi_i(x,t)
\psi_j^*(x,t) \, . \label{e12}
\end{equation}

Qualitatively, the time evolution in Eq.~(\ref{e11}) describes
propagation of the initial wavepacket (\ref{e8}) towards the
scattering potential and then separation of this wavepacket in
coordinate space into the transmitted and reflected parts that are
well-localized on the opposite sides of the scattering region. If
we assume that the scattering potential $U(x)$ has simple shape
(for instance, does not have narrow quasi-bound states) and is
non-vanishing only in some small region of size $\eta \ll l$, the
time $t_{sc}$ from the fluxon injection to completion of the
scattering process is not drastically different from the time
$l/u$ of free fluxon propagation through the JTL. Then at time
$t>t_{sc}$, the separated wavepackets move in the region free from
the $j$-dependent scattering potential and the unitarity of the
quantum-mechanical evolution of $\psi_j(x,t)$ implies that the
overlap of the fluxon wavefunctions in Eq.~(\ref{e12}) becomes
independent of $t$.

This overlap can be directly found in the momentum representation:
\begin{equation}
\int dx \psi_i(x,t) \psi_j^*(x,t)= \int dk |b(k)|^2[t_it_j^*+r_i
r_j^*]\, . \label{e13}
\end{equation}
Here $b(k)$ is the probability amplitude for the fluxon to have
momentum $k$ in the initial state (\ref{e8}), e.g., in the case of
the Gaussian wavepacket (\ref{e10}),
\begin{equation}
b(k) = (\xi^2/\pi)^{-1/4} e^{-(k-k_0)^2\xi^2/2-i(k-k_0)\bar{x}}\,
. \label{e14}
\end{equation}

Equations (\ref{e12}) and (\ref{e13}) show that the diagonal
elements of the density matrix $\rho$ do not change in the process
of scattering of one fluxon, while the off-diagonal elements are
suppressed by the factor
\begin{equation}
\nu = \left|\int dk |b(k)|^2 [t_i(k)t_j^*(k)+ r_i(k)r_j^*(k)]
\right|\leq 1 \, .\label{e14*}
\end{equation}
The inequality in this relation can be proven as the Swartz
inequality for the scalar product in the Hilbert space of
``vectors'' $\{t_j(k),r_j(k)\}$ weighted by $|b(k)|^2$. The
suppression of the off-diagonal elements of $\rho$ due to the
interaction with fluxons is manifestation of the back-action
dephasing of the measured system by the JTL detector. If we add
the suppression factors for the fluxons injected into the JTL with
frequency $f$, the rate of this dephasing is obtained as
\begin{equation}
\Gamma_{ij} = - f \ln \left|\int dk |b(k)|^2 [t_i(k)t_j^*(k)+
r_i(k)r_j^*(k)] \right| \, . \label{e15}
\end{equation}

Equation (\ref{e15}) is similar, but not identical, to the
back-action dephasing rate by the QPC detector \cite{b12}. The
main difference with the QPC is the stage at which the summation
over the momentum $k$ is carried out. This difference reflects the
fact that in contrast to electrons in the QPC, different
$k$-components of the fluxon wavefunction do not scatter
independently. They are constrained by the condition that one
fluxon as a whole is either transmitted or reflected by the
potential, since scattering events of different fluxons are
well-separated in time. As we will see below, this difference
makes it possible to operate the JTL detector in the new,
time-delay mode that is not possible with the dc-biased QPC
detector.

As examples of application of Eq.~(\ref{e15}) we consider several
specific cases motivated by the measurement regimes discussed
below. In one, the phases of the scattering amplitudes are assumed
to cancel out from Eq.~(\ref{e15}), while the variation of the
absolute values of the amplitudes with index $j$ is small. The
dephasing rate (\ref{e15}) can be expressed then in terms of the
variations $\delta T_j(k)$ of the fluxon transmission probability
in different states $|j\rangle$ around some average transmission
$T(k)$: $|t_j(k)|^2=\delta T_j(k)+T(k)$, $\delta T_j(k)\ll T(k)$.
In the lowest non-vanishing order in $\delta T_j(k)$ we get:
\begin{equation}
\Gamma_{ij} = f \int dk |b(k)|^2 \frac{(\delta T_i(k)- \delta
T_j(k))^2}{8T(k)(1-T(k))} \, . \label{e151}
\end{equation}
This equation describes the ``linear-response'' regime of
operation of the JTL detector, when its properties follow from the
general theory of linear detectors \cite{m1}. In particular, the
dephasing (\ref{e151}) can be understood as being caused by the
back-action noise arising from the randomness of the fluxon
transmission/relection process.

In the ``tunnel'' limit of weak transmission, $|t_j(k)|\ll 1$, and
again assuming that the phases of the scattering amplitudes cancel
out, Eq.~(\ref{e15}) reduces to:
\begin{equation}
\Gamma_{ij} = (f/2) \int dk |b(k)|^2 (|t_i(k)|-|t_j(k)|)^2 \, .
\label{e152}
\end{equation}
Both Eqs.~(\ref{e151}) and (\ref{e152}) have direct analogues in
the case of the QPC detectors \cite{b17,b18,b19}.

As the last example that does not have the analogue in the QPC
physics, we consider the situation when the reflection amplitudes
are negligible, $r_j(k)\equiv 0$, and the system-JTL interaction
modifies only the phases $\chi_j(k)$ of the transmission
amplitudes: $t_j(k)=e^{i\chi_j(k)}$, so that
\begin{equation}
\Gamma_{ij} = - f \ln \left|\int dk |b(k)|^2 e^{i[\chi_i(k)-
\chi_j(k)]} \right| \, . \label{e157}
\end{equation}
In coordinate representation, this means that the scattering
potential affects only the position and, in general, the shape of
the transmitted wavepacket. If the width $\delta k$ of the initial
fluxon state in the momentum representation is sufficiently
narrow, and the phases $\chi_j(k)$ do not vary strongly over this
momentum range, they can be approximated as
\begin{equation}
\chi_j(k)= \chi_j(k_0)-(k-k_0)x_j\, , \;\;\; x_j \equiv
-\chi_j'(k_0) \, . \label{e153}
\end{equation}
One can see directly that this approximation neglects distortion
of the fluxon wavepacket in the scattering process, while taking
into account its shift $x_j$ along the coordinate axis. This shift
can be directly related to the ``time delay'' $\tau_j$ due to
scattering: $\tau_j=x_j/u$ (which can be both negative and
positive depending on the scattering potential). The dephasing
rate (\ref{e157}) can then be conveniently written down in the
coordinate representation in terms of the initial fluxon
wavepacket $\psi_0(x)$:
\begin{equation}
\Gamma_{ij} = - f \ln \left|\int dx \psi_0(x-x_i)\psi_0^*(x-x_j)
\right| \, . \label{e154}
\end{equation}
Equation (\ref{e154}) shows explicitly that the back-action
dephasing by the JTL detector arises from the entanglement between
the measured system and the scattered fluxons which are shifted in
time by an interval dependent on the state of the system. The
degree of suppression of coherence between the different states of
the measured system is determined then by the magnitude of the
relative shift of the fluxon in these states on the scale of the
wavepacket width. For instance, if the initial fluxon wavepacket
is Gaussian (\ref{e10}), Eq.~(\ref{e154}) gives:
\begin{equation}
\Gamma_{ij} = f(x_i-x_j)^2/4\xi^2 \, . \label{e155}
\end{equation}

Back-action dephasing represents only part of the measurement
process. The other part is information acquisition by the detector
about the state of the measured system. In the case of the JTL
detector, this information is contained in the scattering
characteristics of fluxons, and the rate of its acquisition
depends on specific characteristics recorded by the fluxon
receiver. There are at least two different possibilities in this
respect. One is to detect the probability of fluxon transmission
through the scattering region (or, equivalently, the corresponding
probability of the fluxon reflection back into the generator, see
Fig.~1). Another possible detection scheme is for the receiver to
measure the time delay associated with the fluxon propagation
through the JTL. Even if the measured system changes the potential
$U_j(x)$ in such a way that the fluxon transmission probability is
not affected, potential can still change the fluxon propagation
time, which will contain then information about the state of the
system. In general, one can have a situation when the information
is contained both in the changes of the propagation time and
transmission probability, and in order not to loose any
information one would need to detect both scattering
characteristics. In this work, we consider only the two ``pure''
cases of transmission and time-delay detection modes.

\subsection{Transmission detection mode}

If the detector records only the fact of the fluxon arrival at the
receiver, then only the modulation of the fluxon transmission
probability by the measured system conveys information about the
system. The information contained in all other features of the
scattering amplitudes (e.g., their phases or propagation time) is
lost in the receiver. In this case, the rate of information
acquisition can be calculated simply by starting with the
probabilities of the fluxon transmission/reflection $T_j$ and
$R_j$, when the measured system is in the state $|j\rangle$:
\begin{equation}
T_j = \int dk |b(k)|^2 |t_j(k)|^2 \, , \;\;\; R_j=1-T_j\, .
\label{e16}
\end{equation}

Since the outcomes of successive fluxon scattering event are
independent, the probability $p(n)$ to have $n$ out of $N$
incident fluxons transmitted, is given by the binomial
distribution $p_j(n)=C_N^n T_j^nR_j^{N-n}$. The task of
distinguishing different states $|j\rangle$ of the measured system
is transformed into distinguishing the probability distributions
$p_j(n)$ for different $j$s. Since the number $N=ft$ of scattering
attempts increases with time $t$, the distributions $p_j(n)$
become peaked successively more strongly around the corresponding
average numbers $T_jN$ of transmitted fluxons. The states with
different probabilities $T_j$ can be distinguished then with
increasing certainty. The rate of increase of this certainty can
be characterized quantitatively by some measure of the overlap of
the distributions $p_j(n)$. While in general there are different
ways to characterize the overlap of different probability
distributions \cite{b20}, the characteristic which is appropriate
in the quantum measurement context \cite{b21,m2} is closely
related to ``fidelity'' in quantum information \cite{b20}: $\sum_n
[p_i(n)p_j(n)]^{1/2}$. The rate of information acquisition
(increase of confidence level in distinguishing states $|i\rangle$
and $|j\rangle$) can then be defined naturally as \cite{b12}:
\begin{equation}
W_{ij} = - (1/t) \ln \sum_n [p_i(n) p_j(n)]^{1/2} \, . \label{e17}
\end{equation}
Using the binomial distribution in this expression we get:
\begin{equation}
W_{ij} = - f \ln [(T_iT_j)^{1/2}+(R_iR_j)^{1/2} ] \, , \label{e18}
\end{equation}
where the transmission and reflection probabilities are given by
Eq.~(\ref{e16}).

Equation (\ref{e18}) characterizes the information acquisition
rate of the JTL detector in the transmission-detection mode. For
an arbitrary detector, the information rate should be smaller than
or equal to the back-action dephasing rate, and regime when the
two rates are equal is ``quantum-limited''. Comparing
Eqs.~(\ref{e15}) and (\ref{e18}) one can see that for the JTL
detector, indeed,
\begin{equation}
W_{ij} \leq \Gamma_{ij} \, , \label{e21}
\end{equation}
and equality holds if several conditions are satisfied. First two
conditions require that there is no information in the phases of
the transmission amplitudes:
\begin{eqnarray}
\phi_j(k) = \phi_i(k) \, , \;\; \phi_j(k) \equiv
\mbox{arg}(t_j(k) /r_j(k)) \, , \label{e19} \\
\chi_j(k)-\chi_i(k)= \mbox{const}\, .  \qquad \qquad
 \label{e191}
\end{eqnarray}
The two conditions, (\ref{e19}) and (\ref{e191}), have different
physical origin. Condition (\ref{e19}) implies that the scattered
states contain no information on $j$ that can be used in principle
by arranging interference between the transmitted and reflected
parts of the wavefunction \cite{b12}. In practical terms, the
simplest way to satisfy this condition is to make the scattering
potential symmetric $U_j(-x)=U_j(x)$ for all states $|j\rangle$.
The unitarity of the scattering matrix for the fluxon scattering
in the JTL implies in this case that $\phi_j =\pi/2$ for any $j$.
Condition (\ref{e191}) means that no information on $j$ is
contained in the shape and position of transmitted wavepackets
that would be lost in the fluxon receiver operating in the
transmission-detection mode. (Similar condition for the reflected
wavepacket follows from Eqs.~(\ref{e19}) and (\ref{e191}).) In
general, condition (\ref{e191}) requires that the spread $\delta
k$ of the initial fluxon state over momentum gives rise to the
uncertainty in the fluxon energy $\delta \epsilon \simeq u \delta
k$ that is much smaller than the energy scale $\Omega$ of the
transparency variation of the scattering potential $U_j(x)$.

One more condition of the quantum-limited operation is that the
fluxon transmission probabilities are effectively momentum- and
energy-independent in the relevant momentum range:
\begin{equation}
|t_j(k)|^2 = T_j \, .  \label{e20}
\end{equation}
This condition again requires that $\delta \epsilon \ll \Omega$.
It is more restrictive than the corresponding condition for the
QPC detector which can be quantum-limited even in the case of the
energy-dependent transmission probability \cite{b22,b23,b12}. To
obtain Eq.~(\ref{e20}), one starts from the back-action dephasing
rate (\ref{e15}) which can be written as
\[ \Gamma_{ij} = - f \ln \int dk |b(k)|^2 [|t_i(k)t_j(k)| +
|r_i(k)r_j(k)|] \] under the conditions (\ref{e19}) and
(\ref{e191}). Swartz inequality for the functions $|t_j(k)|$:
\[ [T_iT_j]^{1/2} \geq \int dk |b(k)|^2 |t_i(k)t_j(k)|\, , \]
and similar inequality for $|r_j(k)|$ show that this $\Gamma_{ij}$
and the information acquisition rate $W_{ij}$ satisfy the
inequality (\ref{e21}). Equality in Eq.~(\ref{e21}) could be
reached only when
\begin{equation}
|t_j(k)|= \lambda_j t(k) \, , \;\;\; |r_j(k)|= \lambda_j' r(k) \,
, \label{e22}
\end{equation}
where $\lambda$'s are some constants, so that the ratios
$|t_i(k)|/|t_j(k)|$ and $|r_i(k)|/|r_j(k)|$ are independent of
$k$. Similarly to Eq.~(\ref{e191}), Eq.~(\ref{e22}) demands that
no information about the state of the measured system is contained
in the shape of transmitted or reflected wavepackets. In general,
when both the transmission and reflection probabilities are not
small, the two relations in (\ref{e22}) are incompatible. They
have only a trivial solution in which all amplitudes are
independent of $k$ in the relevant range of $k$'s, thus proving
Eq.~(\ref{e20}) for $T_j \sim R_j \sim 1/2$. The transmission and
reflection probabilities have roughly the same magnitude when the
fluxon energy $\epsilon$ is close to the maximum $U$ of the
scattering potential. In this case, small spread of $\epsilon$:
$\delta \epsilon \ll \Omega$ implies that the range $\eta$ of the
scattering potential $U(x)$ should be small: $\eta \ll \xi$.

Condition (\ref{e20}) of the quantum-limited operation of the JTL
detector is not necessary when either $T_j\ll 1$ or $R_j \ll 1$.
In this case, one of the relations in (\ref{e22}) reduces to a
trivial statement $|r_j(k)|\simeq 1$ or $|t_j(k)|\simeq 1$. The
other relation gives then the actual condition of the
quantum-limited operation that can be satisfied in principle with
an arbitrary $k$-dependent function $t(k)$ or $r(k)$.

Finally, we show how Eq.~(\ref{e20}) applies in the
linear-response regime, when the variations $\delta T_j$ of JTL
transparencies between the different states $|j\rangle$ are small
and back-action dephasing is given by Eq.~(\ref{e151}). Expanding
Eq.~(\ref{e18}) in $\delta T_j$: $T_j=T+\delta T_j$, where all
transparencies are defines as in Eq.~(\ref{e16}), we get:
\begin{equation}
W_{ij} = f (\delta T_i- \delta T_j)^2/[8T(1-T))]\, . \label{e25}
\end{equation}
This equation differs from Eq.~(\ref{e151}) only by the order in
which the integration over momentum is performed.This means that
the information and dephasing rates satisfy the inequality
(\ref{e21}) and are equal only if the transparency is constant in
the relevant momentum range.

In the linear-response regime, each individual fluxon carries only
small amount of information, and it is convenient to employ the
quasi-continuous description in which the fluxon receiver acts as
the voltmeter registering not the individual fluxons, but the rate
of arrival of many fluxons, i.e., the voltage $V(t)$ across the
junctions of the JTL. The average voltage in the state $|j\rangle$
is
\begin{equation}
\langle V(t) \rangle = f T_j \Phi_0 \, , \label{e23}
\end{equation}
where $\langle ... \rangle$ implies average over the scattering
outcomes and over time $t$ within the fluxon injection cycle.
Because of the randomness of the fluxon scattering, the actual
voltage fluctuates around the average values (\ref{e23}) even at
low frequencies. The voltage fluctuations can be described as the
shot noise of fluxons and its spectral density
\begin{equation}
S_V(\omega ) = \int d\tau e^{-i \omega \tau} (\langle V(t+\tau)
V(t) \rangle- \langle V (t) \rangle^2) \label{e23*}
\end{equation}
is constant at frequencies $\omega$ below the fluxon injection
frequency $f$. Straightforward calculation similar to that for the
regular shot noise shows that this constant is
\begin{equation}
S_0 = f T(1-T) \Phi_0^2 \, , \label{e24}
\end{equation}
where in the linear-response regime we can neglect small
differences $\delta T_j$ of transparencies between the different
states $|j\rangle$ in the expression for noise. This equation
shows that in accordance with the general theory of linear quantum
measurements, the information rate (\ref{e25}) can be interpreted
as the rate with which one can distinguish dc voltage values
(\ref{e23}) in the presence of white noise with the spectral
density (\ref{e24}) \cite{m1,m2}.

To summarize this subsection, we see that in the most relevant
regime $T_j \simeq 1/2$ which maximizes the detector response to
the input signal, conditions of the quantum-limited operation of
the JTL detector are given by Eqs.~(\ref{e19}) and (\ref{e20}).
These conditions are satisfied if the scattering potential for the
fluxons created by the input signal is symmetric and has the range
$\eta$ smaller that the size $\xi$ of the fluxon wavepacket.

\subsection{Time-delay detection mode}

Since the range $\eta$ of the scattering potential (\ref{e6}) can
not be smaller than the fluxon size $\lambda_J$, it might be
difficult in practice to realize the condition $\eta \ll \xi$
needed for the quantum-limited operation of the JTL detector in
the transmission-detection mode. For ``quasiclassical'' potential
barriers that are smooth on the scale of fluxon wavepacket, $\eta
\gtrsim \xi$, the ``transition'' region of energies near the top
of the barrier where the reflection and transmission amplitudes
have comparable magnitude, is narrow. If the interval $\delta
\epsilon$ of the fluxon energies avoids this narrow region, then
either transmission or reflection coefficient can be neglected.
Ballistic motion of fluxons in this regime still contains
information about the potential $U_j(x)$ that can be used for
measurement. This information is contained in the time shift
$\tau_j$ caused by the propagation through the region of
non-vanishing potential. Quantum-mechanically, the time-shift
information is contained in the phases of the scattering
amplitudes. To be specific, we discuss here the regime when the
JTL detector is operated in this ``time-delay'' detection mode
using the transmitted fluxons, i.e., $|t_j(k)|=1$. In the energy
range where $|r_j(k)|=1$, the same detection process is possible
using the reflected fluxons, the advantage of the transmission
case being the possibility to make use of the full range of values
of the scattering potential: $U_j(x)<0$ and $U_j(x)>0$.

For sufficiently smooth potential $U_j(x)$, the phase $\chi_j(k)$
of the transmission amplitude can be calculated in the
quasiclassical approximation:
\begin{equation}
\chi_j(k) = \int dx [2m(\epsilon -U_j(x))]^{1/2} \, . \label{e27}
\end{equation}
If the potential is weak, $U_j(x)\ll \epsilon$, potential-induced
contribution to the phase (\ref{e27}) is
\begin{equation}
\chi_j(k) = -\frac{1}{u} \int dx U_j(x) \, . \label{e28}
\end{equation}

Under the adopted assumption of quasiclassical potential and
vanishing reflection, condition (\ref{e9}) of the negligible
broadening of the fluxon wavepacket still works in the presence of
potential. In this case, one can use approximation (\ref{e153})
for the phases $\chi_j(x)$ which implies the shift of the
wavepacket as a whole without distortion. The potential-induced
part of the shift $x_j=-\chi'_j(k_0)$ follows from Eq.~(\ref{e27})
and has the classical form:
\begin{equation}
x_j = \int dx[1-\frac{u}{u_j(x)}]\, , \;\; u_j(x)= [2(\epsilon
-U_j(x))/m]^{1/2} \, . \label{e29}
\end{equation}
For weak potential, $U_j(x)\ll \epsilon$:
\begin{equation}
x_j = - \frac{1}{2\epsilon} \int dx U_j(x) \, .  \label{e30}
\end{equation}

Back-action dephasing rate by the JTL detector in this regime is
given by Eqs.~(\ref{e154}) and (\ref{e155}). The information about
the states $|j\rangle$ contained in the shift $\tau_j$ of the
fluxon in time or, equivalently, coordinate $x_j=\tau_j u$, can be
read off by distinguishing different shifts $x_j$ against the
background of the finite width $\xi$ of the fluxon wavepacket
$\psi_0(x)$. Since $|\psi_0(x)|^2$ gives the probability of
finding the fluxon at coordinate $x$, this task is equivalent to
the task of distinguishing two shifted probability distributions
(see Fig.~4) that was discussed above for the
transmission-detection mode. Similarly to Eq.~(\ref{e17}), we can
write the information acquisition rate of the JTL detector in the
time-delay mode as follows:
\begin{equation}
W_{ij} = - f \ln \int dx |\psi_0(x-x_i)\psi_0(x-x_j)| \, .
\label{e31}
\end{equation}
Comparing this to the dephasing rate (\ref{e154}), we see that in
general the two rates satisfy the inequality (\ref{e21}) as they
should. The rates are equal if the phase of the initial wavepacket
$\psi_0(x)$ of the injected fluxon is independent of $x$, i.e., if
$\psi_0(x)$ is essentially real. In particular, in the case of the
Gaussian wavepacket (\ref{e14}), the JTL detector is
quantum-limited, $W_{ij}=\Gamma_{ij}$, and the two rates are given
by Eq.~(\ref{e155}). These considerations also imply that the JTL
detector in the time-delay mode would lose the property of being
quantum-limited if the fluxon wavepacket spreads noticeably during
propagation through the JTL. This process creates non-trivial
$x$-dependent phase of the wavepacket and makes the information
acquisition rate smaller than the back-action dephasing rate.

\section{Conditional evolution}

As is the case with any dephasing, the back-action dephasing by
the JTL detector can be viewed as the loss of information. In the
regime of the quantum-limited detection, the overall evolution of
the detector and the measured system is quantum-coherent and the
only source of the information loss is averaging over the
detector. For a detector, different outcomes of the evolution are,
however, classically distinguishable, and it is meaningful to ask
how the measured system evolves for a given detector output. In
the quantum-limited regime, specifying definite detector output
eliminates all losses of information, and as a result there is no
back-action dephasing present in the dynamics of the measured
system conditioned on specific detector output.\\

\begin{figure}[h]
\setlength{\unitlength}{1.0in}
\begin{picture}(3.1,0.9)
\put(.2,-.15){\epsfxsize2.8in\epsfbox{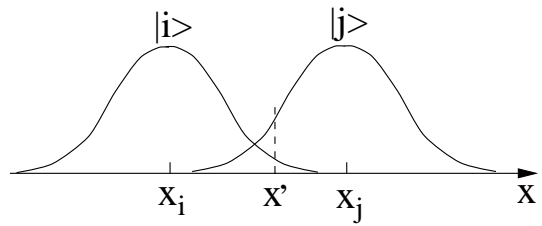}}
\end{picture}
\caption{Illustration of the information acquisition process by
the JTL detector in the time-delay mode. The fluxon wave-packet is
shifted by the distance $x_j$ dependent on the state $|j\rangle$
of the measured system. In conditional description, observation of
the fluxon with position $x'$ changes the system wavefunction
according to \protect Eq.~(\ref{e35}). }
\end{figure}

Conditional description in the quantitative form is obtained (see,
e.g., \cite{m2,b24,b25}) by separating in the total wavefunction
the terms that correspond to a specific classical outcome of
measurement and renormalizing this part of the wavefunction so
that it corresponds to the total probability of 1. In the case of
the JTL detector in the {\em transmission-detection mode}, for
each injected fluxon there are two classically different outcomes
of scattering: transmission and reflection of the fluxon.
Accordingly, the wavefunction of the measured system should be
conditioned on the observation of either transmitted or reflected
fluxon in each cycle of fluxon injection. The evolution of the
total wavefunction ``detector+measured system'' during the
scattering of one fluxon is described by Eq.~(\ref{e11}). If
conditions (\ref{e191}) and (\ref{e20}) of the quantum-limited
detection are satisfied, the transmission and reflection
amplitudes are effectively momentum independent \cite{rem2}.
Coordinate dependence of the scattered fluxon wavepakets is then
the same in different states $|j\rangle$, and can be factored out
from the total wavefunction. The evolution of the measured system
can then be conditioned on the transmission/reflection of a fluxon
simply by keeping in Eq.~(\ref{e11}) the terms that correspond to
the actual outcome of scattering. If the fluxon is transmitted
through the scattered region or reflected from it in a given
injection cycle, amplitudes $c_j$ for the system to be in the
state $|j\rangle$ change, respectively, as follows:
\begin{equation}
c_j \rightarrow \frac{t_j c_j}{\left[\Sigma_j |c_j
t_j|^2\right]^{1/2}}, \;\; c_j \rightarrow \frac{r_j
c_j}{\left[\Sigma_j |c_j r_j|^2\right]^{1/2}} . \label{e32}
\end{equation}
It is important to stress that the changes in the coefficients
$c_j$ for a system with vanishing Hamiltonian (as we assumed from
the very beginning) is unusual from the point of view of
Schr\"{o}dinger equation, and provides quantitative expression of
reduction of the wavefunction in the measurement process.
Equations (\ref{e32}) are similar to those obtained in the
``bayesian'' approach \cite{m2} for the QPC detector in the tunnel
limit.

If dynamics of the JTL detector is not quantum-limited, then
information is lost and dephasing is non-vanishing even in
conditional evolution. To generalize Eqs.~(\ref{e32}) to this case
of finite ``residual'' dephasing, we need to start with
Eq.~(\ref{e12}) for the change of the density matrix $\rho_{ij}$
of the measured system due to the scattering of one fluxon. To
condition this change on the specific outcome of scattering, we
limit the trace over the fluxon coordinate to the range containing
only transmitted or reflected fluxons. In this way we see that
$\rho_{ij}$ changes as
\begin{eqnarray}
\rho_{ij} \rightarrow \rho_{ij} \int dk |b(k)|^2(t_i(k)t_j^*(k))/
\Sigma_j \rho_{jj} T_j \, , \label{e33} \\
\rho_{ij} \rightarrow \rho_{ij} \int dk |b(k)|^2(r_i(k)r_j^*(k))/
\Sigma_j \rho_{jj} R_j \, , \label{e34}
\end{eqnarray}
if the fluxon is, respectively, transmitted or reflected in a
given cycle. Using the discussion of the transmission-detection
mode in Sec.~III, one can see immediately that the requirements
(\ref{e191}) and (\ref{e20}) of the JTL detector being
quantum-limited, are equivalent to the density matrix $\rho_{ij}$
remaining pure in the conditional evolution described by
Eqs.~(\ref{e33}) and (\ref{e34}).

The above derivation of conditional evolution equations can be
repeated with only minor modifications in the {\em time-delay
mode} of the detector operation. In this case, different classical
outcomes of measurements are the observed instances of time when
the fluxon reaches the receiver, that for convenience can be
directly translated into different fluxon positions $x$ at some
fixed time. If the fluxon is observed at the position $x'$ in a
given injection cycle (Fig.~4), the evolution of the amplitudes
$c_j$ of the measured system due to interaction with this fluxon
is:
\begin{equation}
c_j \rightarrow \psi_0(x'-x_j) c_j/\left[\Sigma_j |c_j
\psi_0(x'-x_j)|^2\right]^{1/2} . \label{e35}
\end{equation}
Qualitatively, and similarly to the conditional evolution in the
transmission mode, the sequence of transformations (\ref{e35})
describes ``weak measurement''. The system wavefunction is
localized gradually in one of the states $|j\rangle$ with
increasing number of the fluxon scattering events which lead to
accumulation of information about  $|j\rangle$. In contrast to the
transmission-detection mode, conditional evolution (\ref{e35})
always remains pure under the simplifying assumptions adopted in
this work for the JTL detector: coherent propagation of fluxons
with fixed wavepacket $\psi_0(x)$ and the same velocity.
Fluctuations in the fluxon generator or finite dissipation in the
JTL would create information losses also in this regime.

\section{Non-quantum-limited detection}

Quantum-limited operation of the JTL detector discussed in the
preceding Sections requires, in essence, quantum-coherent dynamics
of fluxons in the JTL. While this dynamics can be observed
experimentally \cite{tun}, the task of realizing it is certainly
very difficult. From this perspective, it is important that
several attractive features of the JTL detector, e.g., large
operating frequency and reduced parasitic dephasing during the
time intervals between the fluxon scattering, remain even in the
non-quantum-limited regime. Although the Josephson junctions of
the JTL and those in the ``external'' parts of the JTL detector
(generator and receiver) can give rise to dissipation and
dephasing not related directly to measurement, in the JTL geometry
(Fig.~1), parasitic dissipation is suppressed due to screening by
the supercurrent flow in the JTL junctions \cite{b11}.

The dominant deviation from the quantum-limited detection should
be associated then with the fluctuations in the fluxon motion.
These fluctuations make the dephasing factor $\nu$ (\ref{e14*})
due to scattering larger than the amount of information conveyed
by the fluxon scattering. Some interesting measurement strategies
are still possible with the JTL non-ideality of this type. The
most natural example is the quantum non-demolition (QND)
measurements of quantum coherent oscillations in a qubit
\cite{b26,b27} which are designed to make the back-action
dephasing by the detector irrelevant. Dynamics of the fluxon
scattering in the JTL detector makes it particularly suitable for
the ``kicked'' version of the QND qubit measurements of a qubit
\cite{b27} or harmonic oscillator \cite{b28}.

Consider, for instance, a qubit with the Hamiltonian
\begin{equation}
H=-(\Delta/2) \sigma_x \, , \label{e40}
\end{equation}
which performs quantum coherent oscillations with frequency
$\Delta$. The qubit is coupled to the JTL through its $\sigma_z$
operator, i.e. the states $|j\rangle$ used above, are the two
eigenstates of $\sigma_z$. We assume that the Hamiltonian
(\ref{e40}) already includes renormalization of parameters due to
the qubit-detector coupling. If the qubit oscillations are weakly
dephased at the rate $\gamma \ll \Delta$ (e.g., by residual
parasitic dissipation in the JTL detector), the time evolution of
the qubit density matrix $\rho$ during the time intervals between
the successive fluxon scattering events can be written in the
$\sigma_z$-representation as follows:
\begin{equation}
\rho(t) = \frac{1}{2} \left[1+e^{-\gamma t}
\left(\begin{array}{cc} x & -i y \\ i y & -x  \end{array} \right)
\right] \, , \label{e41}
\end{equation}
\begin{equation}
\dot{r} =- i \Delta r \, , \;\;\;\; r=x+iy \, , \label{e42}
\end{equation}
where $r(t=0)=\pm 1$ depending on whether the qubit starts at
$t=0$ from the $\sigma_z=1$ or $\sigma_z=-1$ state.

\begin{figure}[h]
\setlength{\unitlength}{1.0in}
\begin{picture}(3.3,1.1)
\put(.3,-.15){\epsfxsize2.7in\epsfbox{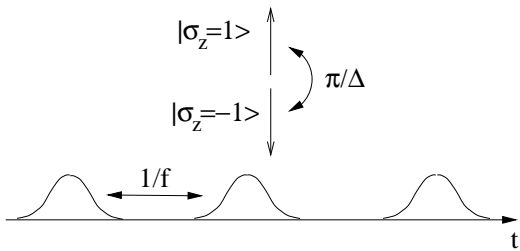}}
\end{picture}
\caption{Schematics of the QND fluxon measurement of a qubit which
suppresses the effect of back-action dephasing on the qubit
oscillations. The fluxon injection frequency $f$ is matched to the
qubit oscillation frequency $\Delta$: $f\simeq \Delta/\pi$, so
that the individual acts of measurement are done when the qubit
density matrix is nearly diagonal in the $\sigma_z$ basis, and the
measurement back-action does not introduce dephasing in the
oscillation dynamics. }
\end{figure}

The fluxon scattering at times $t_n = n/f$ leads to partial
suppression of the off-diagonal elements of $\rho$:
\begin{equation}
y(t_n+0)=\nu y(t_n-0)\, . \label{e43}
\end{equation}
However, if the fluxons are injected in the JTL with the time
interval $1/f$ close to the half-period $\pi/\Delta$ of the qubit
oscillations (Fig.~5), the qubit density matrix (\ref{e42}) is
nearly diagonal at the moments of scattering, $y(t_n)\ll 1$, and
suppression (\ref{e43}) does not affect the qubit strongly. Such a
QND measurement is possible with the JTL detector operating in any
detection mode; to be specific we assume the transmission mode.
The dependence of the fluxon transmission probability on the qubit
state can be written then as $T+\sigma_z \delta T$, where $\delta
T \ll T$ in the ``linear-response regime''. For quantum-limited
detection, the linear-response condition $\delta T \ll T$ implies
that $\nu \rightarrow 1$. In the non-quantum-limited case, the
back-action can be stronger, and we take $\nu$ to be arbitrary
within the $[0,1]$ interval.

If the frequency $f$ is matched precisely to the qubit
oscillations, $f=\Delta/\pi$, the detector does not affect the
qubit dynamics  at all. If the mismatch is non-vanishing but
small, $\delta \equiv \Delta/f -\pi \ll 1$, diagonal elements of
$\rho$ (\ref{e42}) evolve quasi-continuously even if suppression
factor $\nu$ is not close to 1. Equations (\ref{e42}) and
(\ref{e43}) give the following equation for this quasi-continuous
evolution:
\begin{equation}
\dot{x} =- [(1+\nu)/(1-\nu)] (f\delta^2/2) x \, . \label{e44}
\end{equation}

In the assumed linear-response regime, the qubit oscillations
manifest themselves as a peak in the spectral density
$S_V(\omega)$ (\ref{e23*}) of the voltage $V$ across the JTL
junctions. For $f \simeq \Delta/\pi$ the oscillation peak in
$S_V(\omega)$ is at zero frequency. Equations (\ref{e41}) and
(\ref{e44}) describing the decay of correlations in the qubit
dynamics in the $\sigma_z$ basis imply that the oscillation peak
has Lorentzian shape:
\begin{equation}
S_V(\omega ) = S_0+ \frac{2\Gamma f^2 (\delta T)^2
\Phi_0^2}{\omega^2+\Gamma^2} \, ,\label{e45}
\end{equation}
and the oscillation linewidth $\Gamma$ is:
\begin{equation}
\Gamma = \gamma+ \frac{1+\nu}{1-\nu} \frac{(\Delta-\pi f)^2}{2f}\,
. \label{e46}
\end{equation}

For the quantum-limited detection, $\nu \rightarrow 1$, and
Eq.~(\ref{e46}) reproduces previous results for the QND
measurement \cite{b27}, if one introduces the back-action
dephasing rate $\Gamma_d=f(1-\nu)$. In this case, Eq.~(\ref{e46})
is valid for sufficiently small mismatch between the measurement
and oscillation frequencies, $|\delta | \ll 1-\nu$. For larger
$\delta$, the oscillation peak in the detector output
$S_V(\omega)$ moves to finite frequency $\Delta-\pi f$ and the QND
nature of the measurement is lost \cite{b26}. Equation (\ref{e46})
shows also that in the limit of ``projective'' measurements
$\nu=0$, the broadening of the oscillation peak is weaker, and the
peak remains at zero frequency for all reasonable values of the
detuning parameter $|\delta|\ll 1$. Therefore, the stronger
back-action of the JTL detector is advantageous for the QND
measurements of coherent oscillations. The last remark is that
although our discussion here assumed that the fluxon arrival times
are spaced exactly by $1/f$, Eqs.~(\ref{e45}) and (\ref{e46})
should remain valid even in the presence of small fluctuations of
the measurement times. These fluctuation can be described by
taking into account that the detuning $|\delta|$ cannot be made
smaller that the relative linewidth of the fluxon generator.

\section{Conclusion}

We have suggested and analyzed a new ballistic detector for the
rapid read-out of flux qubits which can be implemented with the
present-day SFQ fabrication technology. The detector should
combine quantum-limited dynamics with time resolution that is
better than the decoherence time of typical flux qubits, making it
possible to perform measurements on quantum-coherent qubits. To
operate in the quantum limited regime, the detector requires
junctions with sub-micron size, and its time resolution in this
regime should reach the range of 0.3 ns. The detector is
potentially useful even in the non-quantum-limited regime, when
the time resolution can be improved further, and the detector can
perform, for instance, the QND measurements of coherent
oscillations.

\section{Acknowledgment}

We would like to thank A. Ustinov for useful discussion.  This
work was supported in part by ARDA and DOD under the DURINT grant
\# F49620-01-1-0439, by the NSF under grant \# DMR-0325551, and by
CREST.


\begin{thebibliography}{99}

\bibitem{b1} J.R. Friedman, V. Patel, W. Chen, S.K. Tolpygo, and J.E.
Lukens, Nature {\bf 406}, 43 (2000).

\bibitem{b2} D. Vion, A. Aassime, A. Cottet, P. Joyez, H. Pothier,
C. Urbina, D. Esteve, and M. H. Devoret, Science {\bf 296}, 886
(2002).

\bibitem{b3} Y. Yu, S.Y. Han, X. Chu, S.I. Chu, and Z. Wang,
Science {\bf 296}, 889 (2002).

\bibitem{b4} I. Chiorescu, Y. Nakamura, C.J.P.M. Harmans, and J.E.
Mooij, Science {\bf 299}, 1869 (2003).

\bibitem{b5} A.J. Berkley, H. Xu, R.C. Ramos, M.A. Gubrud,
F.W. Strauch, P.R. Johnson, J.R. Anderson, A.J. Dragt, C.J. Lobb,
and F.C. Wellstood, Science {\bf 300}, 1548 (2003)

\bibitem{b6} S. Saito, M. Thorwart, H. Tanaka, M. Ueda, H. Nakano,
K. Semba, and H. Takayanagi, Phys.\ Rev.\ Lett. {\bf 93}, 037001
(2004).

\bibitem{b7} J. Claudon, F. Balestro, F. W. J. Hekking, and O. Buisson
Phys.\ Rev.\ Lett. {\bf 93}, 187003 (2004).

\bibitem{b8} R. McDermott, R. W. Simmonds, M. Steffen, K.B. Cooper, K.
Cicak, K.D. Osborn, S. Oh, D.P. Pappas, and J.M. Martinis, Science
{\bf 307} 1299 (2005).

\bibitem{b9} B.L.T. Plourde, T.L. Robertson, P.A.
Reichardt, T. Hime, S. Linzen, C.-E. Wu, J. Clarke,
cond-mat/0501679.

\bibitem{rem1} Another way of avoiding the transition into the dissipative
state in the course of measurement is to detect the variations in
the junction impedance induced by the measured system.  Such
``impedance-measurement'' schemes \cite{b10*,b11*} enable one to
perform the measurements continuously, but they typically require
narrow-band coupling between the detector and the measured system,
the fact that limits their time resolution.

\bibitem{b10*} E. Il'ichev, N. Oukhanski, A. Izmalkov, Th. Wagner, M.
Grajcar, H.-G. Meyer, A.Yu. Smirnov, A. M. van den Brink, M.H.S.
Amin, and A. M. Zagoskin, Phys.\ Rev.\ Lett. {\bf 91}, 097906
(2003).

\bibitem{b11*} A. Wallraff, D.I. Schuster, A. Blais, L. Frunzio, J.
Majer, M.H. Devoret, S.M. Girvin, and R.J. Schoelkopf, Phys.\
Rev.\ Lett. {\bf 95}, 060501 (2005).

\bibitem{b10} K.K. Likharev and V.K. Semenov, IEEE Trans.\ Appl.\
Supercond. {\bf 1}, 3 (1991).

\bibitem{b11} V.K. Semenov and D.V. Averin, IEEE Trans. Appl.
Supercond. {\bf 13}, 960 (2003).

\bibitem{b12*} V.K. Kaplunenko and A.V. Ustinov, Eur.\ Phys.\ J.\
B {\bf 38}, 3 (2004).

\bibitem{b13*} A.M. Savin, J.P. Pekola, D.V. Averin, and V.K. Semenov,
cond-mat/0509318.

\bibitem{b12} D.V. Averin and E.V. Sukhorukov, Phys.\ Rev.\ Lett.
{\bf 95}, 126803 (2005).

\bibitem{sol} R. Rajaraman, "Solitons and Instantons", (North Holland,
N.Y., 1982).

\bibitem{b13} A. Schmid, Phys.\ Rev.\ Lett. {\bf 51}, 1506 (1983).

\bibitem{b14} D.V. Averin, A.B. Zorin, and K.K. Likharev,
Sov.\ Phys.\ JETP  {\bf 61}, 407 (1985).

\bibitem{b15} Yu. Koval, A. Wallraff, M.V. Fistul, N. Thyssen,
H. Kohlstedt, and A.V. Ustinov, IEEE Trans.\ Appl.\ Supercond.
{\bf 9}, 3957 (1999).

\bibitem{b16} A. Shnirman, E. Ben-Jacob, and B. Malomed, Phys.\
Rev.\ B {\bf 56}, 14677 (1997).

\bibitem{p1} M.B. Fogel, S.E. Trullinger, A.R. Bishop, and J.A.
Krumhansl, Phys.\ Rev.\ Lett. {\bf 36}, 1411 (1976).

\bibitem{p2} D.W. McLaughlin and A.C. Scott, Phys.\ Rev. A {\bf 18},
1652 (1978).

\bibitem{tun} A. Wallraff, A. Lukashenko, J. Lisenfeld, A. Kemp,
M.V. Fistul, Y. Koval, A.V. Ustinov, Nature {\bf 425}, 155 (2003).

\bibitem{dis} O.M. Braun and Y.S. Kivshar, ``The Frenkel-Kontorova
model'' (Spinger, Berlin, 2004).

\bibitem{m1} D.V. Averin, in: "Quantum Noise in Mesoscopic Physics",
Ed. by Yu.V. Nazarov, (Kluwer, 2003) p. 229; cond-mat/0301524.

\bibitem{b17} I.L. Aleiner, N.S. Wingreen, and Y. Meir, Phys.\
Rev.\ Lett. {\bf 79}, 3740 (1997).

\bibitem{b18} Y. Levinson, Europhys.\ Lett. {\bf 39}, 299 (1997).

\bibitem{b19} S.A. Gurvitz, Phys.\ Rev. B {\bf 56}, 15215 (1997).

\bibitem{b20} M.A. Nielsen and I.L. Chuang, {\em ``Quantum computation
and quantum information''}, (Cambridge, 2000), Ch.~9.

\bibitem{b21} W.K. Wootters, Phys.\ Rev. D {\bf 23}, 357 (1981).

\bibitem{m2} A.N. Korotkov, in: "Quantum Noise in Mesoscopic Physics",
Ed. by Yu.V. Nazarov, (Kluwer, 2003) p. 205.

\bibitem{b22} S. Pilgram and M. B\"{u}ttiker, Phys.\ Rev.\ Lett.
{\bf 89}, 200401 (2002).

\bibitem{b23} A.A. Clerk, S.M. Girvin, and A.D. Stone, Phys.\
Rev. B {\bf 67}, 165324 (2003).

\bibitem{b24} J. Dalibard, Y. Castin, and K. M\o lmer, Phys.\
Rev.\ Lett. {\bf 68}, 580 (1992).

\bibitem{b25} H.-P. Breuer and F. Petruccione, {\em The theory of
open quantum systems}, (Oxford Univ. Press, 2002).

\bibitem{rem2} Condition (\ref{e19}) of the quantum-limited
operation arises from mixing of the transmitted and reflected
parts of the wavefunction and does not directly play a role in the
conditional evolution.

\bibitem{b26} D.V. Averin, Phys.\ Rev.\ Lett. {\bf 88}, 207901 (2002).

\bibitem{b27} A.N. Jordan and M. Buttiker, Phys.\ Rev. B {\bf
71}, 125333 (2005).

\bibitem{b28} R. Ruskov, K. Schwab, and A.N. Korotkov, Phys.\ Rev. B {\bf
71}, 235407 (2005).


\end{thebibliography}
\end{document}